# Félix de Roy: a life of variable stars

Jeremy Shears

## Abstract

Félix de Roy (1883-1942), an internationality recognised amateur astronomer, made significant contributions to variable star research. As an active observer, he made some 91,000 visual estimates of a number of different variable stars. A Belgian national, he took refuge in England during World War 1. While there, de Roy became well enough known to later serve as Director of the BAA Variable Star Section for seventeen years. Through this office, and his connections with other organisations around the world, he encouraged others to pursue the observation of variable stars. Not merely content to accumulate observational data, de Roy also analysed the data and published numerous papers.

## Introduction

I first became aware of the variable star work of Félix de Roy when I was researching historical observations of HR Lyr (Nova Lyrae 1919) as it turned out that he had observed this star and I included his observations in a subsequent paper (1). I was intrigued to learn that de Roy served as Director of BAA Variable Star Section (BAA-VSS) between 1922 and 1939 and that he fulfilled this role whilst living in Belgium. Not only did this strike me as being unusual for the time, but as I have personal links with, and a high level of respect for, the Belgian amateur astronomical community, having cooperated with several of them in variable star research, I became intrigued. Moreover, Belgium was my home for 5 very happy years. A little research revealed that de Roy was very active in both observational astronomy, especially variable stars and meteors, as well as in the analysis of variable star observations. He was also involved in popularising and organising observational astronomy through his involvement with a number of national and international astronomical associations. Surprisingly, rather little has been written about de Roy's remarkable life. A detailed obituary was published in French (2), which was drawn on in preparing this paper, and a brief English obituary followed (3). More recently Thomas Williams presented a summary of his life at the AAVSO meeting held in Belgium in 1990 (4). Thus the aim of this paper is to highlight the important contribution made by de Roy to variable star research and astronomy in general. A portrait of de Roy is shown in Figure 1.

## Education and professional career

Félix Eugène Marie de Roy was born in the Borgerhout district of Antwerp (5), Belgium, on the 25[th] July 1883 (a map of Belgium is shown in Figure 2). He went to college at the Athenaeum in Ixelles, a district of Brussels (6), where he excelled academically. He showed a strong interest in science and mathematics, conducting scientific experiments in his spare time. The need to earn a living took him into journalism as a career, writing at different times for the newspapers La Métropole, l'Action National, Le Neptune and Le Matin (7). The titles for which he wrote were of different social and political leanings, but de Roy's contributions were known for their objectivity and balance. His professional integrity was widely admired, leading his colleagues to elect him as President of the Antwerp-Limburg section of the General Association of the Belgian Press (2).

**The Astronomical Society of Antwerp**

De Roy was one of a small group of amateur astronomers who in 1905 founded the Astronomical Society of Antwerp (Société d'Astronomie d'Anvers, Koninklijk Sterrenkundig Genootschap van Antwerpen), which became the Royal Astronomical Society of Antwerp and is still active today. By the time of the outbreak of the First World War, the Society had about 300 members (thus was almost one-third the size of the BAA), of which 60 to 70 were active observers (8). Initially de Roy was the Society's Secretary, becoming Secretary General in 1911 following the death of Frans Gittens, and was ultimately elected as President on 22nd January 1937. From the beginning he was actively involved in the work and organisation of the young Society and presented numerous lectures on a wide range of astronomical subjects. Many of these were aimed at encouraging members to take up observational astronomy and his lecture on 19th December 1912 (3) on the role played by amateur astronomers in variable star research was typical of his desire to bring others into this area. It was mainly through de Roy's initiative that in 1908 the Society started the Gazette Astronomique, a monthly bulletin reporting the activities of the Society and its members. The Society had a small observatory on the roof of a warehouse in the city docks in Antwerp, containing a 4 inch (10 cm) (9) Bardou refractor, a 10 inch (25 cm) reflector by Vincart and an unsilvered 6 inch (15 cm) reflector for solar work (8) (10). Accessories used on the refractor included two spectroscopes—a prominence type and a stellar of the Secchi type, by Zeiss. However, local air pollution meant that observing was challenging. De Roy and other members of the Society observed the partial solar eclipse of 28$^{th}$ June 1908 from the observatory (11).

De Roy helped to organise a successful expedition by 15 members of the Astronomical Society of Antwerp to observe the solar eclipse on 17$^{th}$ April 1912 (12). They set up their observing station on the line of totality near the village of Silenrieux, in the province of Namur (13) in southern Belgium (the complete path of the eclipse is shown in Figure 3). This was an unusual *hybrid eclipse*, or *annular/total eclipse*, which was seen as an annular eclipse along parts of the track and as total on others. De Roy and his team observed what he described as an intermediate event, neither complete totality nor obvious annularity, with the moon not quite covering the diameter of the sun. Whilst this meant that they did not see the chromosphere or prominences, they did observe the corona. Many photographs were taken and de Roy's report of the expedition also notes the observation of shadow bands which appeared 20 seconds before totality, were invisible during totality and reappeared following totality. These were moving across the ground at about 3 to 4 metres per second, being 2 to 4 centimetres wide and 7 to 20 cm apart.

**The Great War, life in England and a Nova**

For centuries, Belgium has been "the Battlefield of Europe", but during the first half of 1914, life carried on pretty much as usual. In June 1914 de Roy represented the Astronomical Society of Antwerp at an international congress of astronomical associations hosted by the Paris Observatory between the 22$^{nd}$ and 25$^{th}$ of that month. An excursion was organised to Camille Flammarion's (1842 – 1925) observatory at Juvisy and a photograph of the participants is shown in Figure 4 (14).

The peace was soon to be shattered. Germany declared war France on 3$^{rd}$ August 1914. The next day Great Britain declared war on Germany and the German army invaded Belgium. As the advance continued, many Belgians fled abroad, in particular to Great Britain. On 20$^{th}$ August the Belgian Army fell back to the "National Redoubt of Antwerp (15) and the Siege of Antwerp began on 28$^{th}$ September. As editor of La Métropole, de Roy kept the presses rolling until the evening of 7$^{th}$ October 1914 (16). He left Antwerp the following day as the bombardment of the city itself (which eventually fell (17)) started, took a ferry across the English Channel and arrived in London. There he resumed producing the newspaper with a break of only 15 days. He recounted some of his experiences in leaving Belgium during the BAA meeting in London on 25$^{th}$ November 1914 (8). Fortunately the Astronomical Society of Antwerp's observatory had been spared by the German shells (10), but he reported that the Royal Observatory (18) in Brussels had fallen into enemy hands, but had not been damaged. Several other Belgian astronomers had also sought refuge in England (and Holland) including several members of the Astronomical Society of Antwerp. De Roy continued publication of the Society's Gazette Astronomique during the war with an editorial board comprising himself and two compatriots, Messrs. J. Linssen (President of the Society) and H. Dierckx (19). Some articles were published in English and were contributed by several well-known British observers, including W.F. Denning (1848-1931), who wanted to show their support for their Belgian friends. At the end of his BAA presentation, de Roy extended his thanks to "the members of the British Astronomical Association, in the name of his brother astronomers, for the kind reception given to them by the President, and also to acknowledge the very warm and cordial hospitality given to the poor destitute Belgians by the English people generally" (20).

BAA meeting reports show that de Roy attended several other meetings in London during the war, as well as at the West of Scotland Branch meeting held on 23$^{rd}$ March 1916 at the Royal Technical College in Edinburgh (21). His subject was "Astronomy in Belgium: Old and New" in which he referred to people with astronomical links, who were associated with what is now Belgium. These included Mercator (1512 – 1594); Simon Stevin (1548/49 (22) – 1620), mathematician, promoter of the decimal system of notation and inventor of the land yacht; Langrenus (1598/1600 (23) – 1675), the creator of modern selenography; Ferdinand Verbiest (1623 – 1688), a Flemish Jesuit missionary to China in the Qing dynasty and director of the Peking Observatory under Emperor Kangxi; Adolphe Quetelet (1796 – 1874), statistician, anthropologist and founder of the Royal Observatory in Belgium; and Jean-Charles Houzeau de Lehaie (1820 – 1888), journalist, Director of the Royal Observatory after Quetelet, who played a role in the American Civil War and who observed the transit of Venus in 1882 in Texas. De Roy also described the instruments and work being conducted at the time of the outbreak of war at the Royal Observatory.

The speaker before de Roy at the November 1914 BAA meeting was the famous double star observer Robert Jonckheere (1889-1974; Figure 5) of the Lille Observatory in France who had also described his flight from France in the face of the enemy in a similar manner to de Roy. Jonckheere recounted how the Germans arrived in the neighbourhood of the Observatory on the 3$^{rd}$ October 1914 and he put his wife and children on the last boat train leaving for London. He remained in Lille for a further six days during the bombardment until the authorities ordered all men to leave the town. Since the trains had been commandeered by the army, he had to walk for three days and nights to Boulogne, when he caught a ferry to England (24). Jonckheere very soon found employment at the Royal Greenwich Observatory, which lasted until the end of the war (24) (25).

Initially based in East Grinstead, de Roy lived for most of the war at 44 Bridport Road, Thornton Heath, near Croydon (now part of London). He was taken under the wing of the well-known amateur astronomer Fiammetta Wilson very early on. Fiammetta Wilson (1864-1920) joined the BAA in 1910 and was especially active in the Meteor Section, teaming up with W.F Denning (26). She evidently had a wide circle of friends, was widely travelled and was a member of several overseas astronomical societies including the Astronomical Society of Antwerp (27) (other BAA members, including Grace Cook, also joined the Society and offered their support to the their Belgian friends stranded in England (28)). She encouraged English astronomers to subscribe to the Gazette Astronomique during the war, a view supported by W.F. Denning in a letter to the English Mechanic (29).Wilson had visited Belgium in the pre-war years and had dinner with de Roy at the Hotel de l'Europe in Antwerp (30). Wilson loaned de Roy a 3.5 inch (8.9 cm) Watson refractor, which he used at x30 (31) or x60 (32), as he had had to leave his 8 inch (20 cm) reflector in Belgium when he escaped. He was obviously most grateful for this kindness and frequently acknowledged in his writings that his observations were made courtesy of the telescope she had lent him. It appears that he occasionally observed from her residence.

With Wilson's telescope to hand, De Roy was able to resume variable star observing in April 1915. In addition, he used 3x43 opera glasses by Steward (31) and the naked eye. During 1915 he made 670 observations of 26 long period variables (LPVs) (32). These include a well-observed maximum of T UMa around 15$^{th}$

September 1915 (32) (Figure 6) and two successive maxima of Mira Ceti (31). He drew attention to the fact that the light curve (Figure 7) showed some unusual features. The first maximum, in January, was unusually faint (v= 3.8): the faintest observed since 1896 and one of the faintest on record at that time. This was followed by an exceptionally bright minimum in August (v=8.7) (33), whereas the next maximum, in December was normal (v=3.0). De Roy observed on slightly fewer nights in 1916 (138) compared with 1915 (150), but he was still able to amass 970 observations of LPVs (34).

One of the most important astronomical events of 1918 was the appearance of Nova Aquilae (now V603 Aql), the brightest nova of the 20$^{th}$ century. De Roy made an independent discovery of the nova on 8$^{th}$ June 1918 at 22.45 UT (35). He described the discovery circumstances in a letter to the English Mechanic:

> "On directing my binoculars on R Scuti as part of my regular variable star work yesterday evening, Saturday, June 8, 10 h.45 m. (G.M.T.), I was immediately struck by the presence to the north of the familiar field of a bright and white star of the first magnitude..... As no discovery of a Nova had appeared in the daily Press, and as I could not ascertain if it had been announced by the Central Bureau of Copenhagen, I thought fit, as early as possible on Sunday morning, to wire the "news" of the night to the Astronomer Royal and to several friends, both here and abroad" (36)

Being such a bright nova, there were many independent discoveries from around the world, including one by Harold Thomson (37) of Newcastle-upon-Tyne who saw it only 1 hour after de Roy (23.44.UT), and which was reported in the same edition of the English Mechanic (38). Other BAA members who made independent discoveries the same evening included Grace Cook (21.30 UT) (39), W.F. Denning (22.00 UT) (40), Mr Packer (22.00) (40), the VSS Director Charles Lewis Brook (22.15 UT) (40), W.H. Steavenson (22.30 UT) (41). However, there were many claims by people around the world that they had seen the nova several days earlier. By comparison with observations from reliable observers and from data by photographic sky patrols, de Roy showed these to be false as in fact the nova had brightened very quickly (42) (43). Thus, as the famous US variable star observer Leslie Peltier (1900-1980) wryly commented in his autobiography *Starlight Nights*, "even in 1918, such things as vivid imaginations, poor memories, and notoriety seekers already were abroad in the land" (44).

De Roy encouraged observers to increase the accuracy of their observations of the nova by publishing and promoting the use of consistent comparison star sequences (45). He also advocated taking into account the effects of differential atmospheric absorption between the nova and the comparisons (46). This was typical of his constant desire to maximise the quality and consistency of variable star observations.

Several observers reported rapid changes in brightness as they watched the star, including Grace Cook, E.E. Markwick, W. Goodacre (1856-1938) and 3 other observers. By contrast, according to C.L. Brook, "Mr Steavenson and others do not believe in the reality of these short period changes" (47). Félix de Roy appeared to doubt the validity of his observation:

> "I don't believe in *real* short period fluctuations of the light of the Nova, but I think the aspect was peculiar. What this peculiarity *is* seems difficult describe; it was something more glowing and piercing that an ordinary star, and quite independent of colour".

It is now know that this phenomenon, known as flickering, occurs in many cataclysmic variables including novae and dwarf novae, and has been observed by both visual and electronic means (48) (49). Given that the phenomenon was observed by at least 6 other observers in Nova Aquilae, it seems likely that it was real.

**Director of the BAA-VSS**

De Roy had begun observing variable stars in earnest in 1906, when in his early 20s. From the start his list of variables comprised all those on the BAA-VSS programme, some variables from the Harvard College Observatory, plus some suspected variables of his own choosing (50) (51). His main instrument throughout his life (with the exception of his stay in England) was an 8 inch (20 cm) reflector made by the Belgian telescope maker Paul Vincart (52), who was later the Vice President of the Astronomical Society of Antwerp and who made the Society's 10 inch (25 cm) mentioned before. De Roy was elected a member of the BAA on 23$^{rd}$ May 1906 (53), having been proposed by two notable amateur astronomers, E.E. Markwick and Julien Péridier (54). Markwick (1853-1925) was then Director of the BAA-VSS (1900-1909). Péridier (1882-1967; Figure 8) was another active BAA-VSS member of French nationality, who later founded a private observatory established in 1933 at Le Houga in the south-west of France comprising an 8 inch (20 cm) visual and photographic f/13 double refractor and a 12 inch (30 cm) Calver reflector (55). After his death, Péridier's library and instruments were acquired by the McDonald Observatory at the University of Texas for use in teaching astronomy.

Towards the end of 1921, C.L. Brook (1855-1939) resigned as Director of the BAA-VSS and de Roy, already very well known both in BAA and in variable star circles, was appointed as the 4$^{th}$ Director by Council (56), starting on 1$^{st}$ January 1922. Since de Roy was living in Belgium, he appointed A.N. Brown as BAA-VSS Secretary to "receive at his address all observations of members of the Section residing in the British Empire", to distribute charts and observation forms, to assist new observers and to archive original observations. Thus with Brown effectively keeping the Section running on a day to day basis, de Roy himself mainly focussed on the equally important work of analysing the data and preparing reports. One of his first duties was to set out his vision for the BAA-VSS, indicating that he expected to make few changes in the programme initially, and, in his usual encouraging style, appealed "to every member of the Association who may be willing to make useful and systematic observations, to join our little band of workers. They may rest assured of the heartiest welcome, and will receive all the assistance at my disposal" (57). Shortly afterwards he and Brown wrote a lengthy introduction and practical guide to the observation of variable stars, which would be useful even to today's visual observer (58).

Within days of becoming Director he issued BAA-VSS Circular number 1 (59), the first of 11 during his Directorship, to members of the Section (the first page is shown in Figure 9) where he listed the stars on the observational programme and proposed a Section meeting (60). Although he numbered the Circulars from 1, Circulars had previously been issued by E.E. Markwick during his Directorship (61). Thus his first Section meeting was held on 25$^{th}$ October 1922 in the Smoking Library at Sion College, London, which was then the venue of main BAA meetings. The meeting was timed so as to take place just before the Association's AGM and lasted 1 hour. De Roy travelled to London for the occasion and 7 other members attended including W.H. Steavenson (1894-1975), the Mars Section Director (62). Further Section meetings were held at Sion College in October 1923 (63) (64) and November 1924 (65), which de Roy was unable to attend, and a fourth in April 1935 (66) . By contrast to today's BAA-VSS meetings, where members and visiting speakers present their results or give talks on the astrophysics of variable stars, the meetings covered more organisational and operational aspects such as Section policy, stars on the Section's programme, how to improve consistency of observations and the philosophy of publishing the Section's results.

The International Astronomical Union (IAU) commission on Variable Stars met in May 1922 in Rome, which de Roy attended. At the meeting the subject of closer co-operation between the various variable star organisations was discussed. Resolutions 3 and 4 suggested that variable star organisations should "enter into friendly correspondence through their Presidents or otherwise, on questions of observation, and discussion and publication of variable star work" and "act in close co-operation" to this effect (67). De Roy, having recently become BAA-VSS Director, recognised that many of the same stars (mainly LPVs) were observed by the various variable star organisations. He was keen to reduce this duplication and suggested that stars should be divided up amongst the organisations, with the further benefit of allowing the organisations to add new stars to their programmes (68). He proposed that several stars (e.g. Mira Ceti, χ Cyg, R Dra, R Aql and S UMa) should be observed by *all* organisations to cross-check the quality of the data and he advocated the use of the same comparison star sequences for these "check" stars. He made a specific proposal along these lines to the American Association of Variable Star Observers (AAVSO), suggesting that the BAA-VSS should retain its existing 48 LPV programme stars (67). The AAVSO Recorder, Leon Campbell (1881-1951) (Figure 10), responded by agreeing that there were indeed some merits to the proposal, but he commented that in practise "it will be difficult to get some of our observers to give up some of their old familiar friends and pets" (69). Nevertheless, Campbell said he would "be pleased to enter into correspondence with you and with other variable star associations" on the proposal. However, nothing further came of it.

The Section flourished under de Roy's Directorship, with membership increasing to thirty active observers and more than 147,495 observations were made (70). He forged links with other variable star groups around the world including the AAVSO. He published regular detailed reports in the JBAA of observations undertaken by members of the Section (Table 1). In each case the observers' names were listed, along with a description of the methods of analysis and summaries of each star. But de Roy's *magnum opus* was the BAA Memoir on the Section's observations of LPVs between 1925 and 1929 (71). This volume contains more than 400 pages and contains 59938 observations of 51 LPVs. It was very well presented and produced, presumably benefitting from de Roy's journalistic and editorial skills.

The BAA-VSS Secretary, A.N. Brown, died in November 1934 and de Roy appointed W.M. ("Max") Lindley (1891-1972) to succeed him as Secretary (72). Over the next few years, Lindley effectively took over the day to day running of the Section (73) and eventually Lindley succeeded de Roy as Director when the latter resigned due to poor health at the time of the outbreak of the Second World War (74).

**The storm clouds of war return**

The 1920s and 1930s were very productive years for de Roy both in his professional life and in astronomy, but this was not to last. Having endured one war, Belgium was sadly to undergo its second tragedy of the 20$^{th}$ century and de Roy makes reference to the gathering storm clouds in his frequent letters to Leon Campbell, AAVSO Recorder, with whom he developed a close and friendly relationship. Campbell had apparently arranged for him to become a member of the AAVSO free of charge in 1921 (75), writing to de Roy "If you wish to become a member of the AAVSO, I am sure we can arrange it so that you would be under no financial obligation. Having such an authority associated with us as a member would be sufficient recompense to us. If you say the word, I shall be pleased to nominate you for membership" (76). In 1928 he was elected to AAVSO Honorary membership (75).

As late as August 1938, de Roy was cheerfully writing to Campbell about the wonderful time he had at the Stockholm meeting of the IAU that month (77). He had enjoyed "[b]eautiful but somewhat hot weather all along, and a cordial reception by the Swedes. Stockholm is a very beautiful city, quite neat and attractive, with its fine water surfaces and beautifully wooded surroundings". He provided Campbell with a summary of the discussions at IAU Commission 27, on Variable Stars, of which he was a member. He also mentions some of the famous astronomers he met there, including Harlow Shapley (1885-1972) Director of Harvard College Observatory, Annie Jump Cannon (1863-1941), W.H. Steavenson and Margaret & Newton Mayall (1902–1996 and 1904 -1989 respectively).

However, by April 1939, de Roy's tone had changed as he gravely commented on the policy of appeasement of the major powers towards the Nazi government in Germany:

> "The situation in Europe is still bad. I do not believe the Germans would be so stupid as to make war, seeing that they can get almost anything by threats, but it cannot be denied that a spark may, at any moment, set fire to the powder barrel, and we are ready for anything.
>
> You are lucky to be out of such a heavy atmosphere, though I see even the USA cannot escape the necessity of increasing their armaments on a tremendous scale. It is a thousand pities that so much good money must be spent on unprofitable work. But let us still hope for the best".

On 1st September 1939, Germany invaded Poland resulting in Britain, France, Australia and New Zealand declaring war on Germany on 3rd September. Meanwhile in de Roy's letter to Campbell in October 1939 he comments that his variable star observations for August and September were rather fewer in number due to the fact that he had moved house, which meant that he had to set up his observing equipment again. Moreover he had been very busy at work, because some of the newspaper staff had had to leave for war duties (78). The international situation was still causing him anxiety:

> "My two sons are in the Belgian Army, the elder.....in the field artillery, the younger in a crack foot regiment. We [i.e. Belgium] have managed to remain neutral, and we trust that Hitler will look twice before adding 700,000 men to his enemies".

De Roy was more optimistic in March 1940 as he sent his monthly observational report to Campbell, saying

> "the situation here looks somewhat better for us, and we hope to keep out of the melting pot, at least during next winter.......Let us hope a change in the inner regime of Germany may bring about the end of the war" (79).

His sense of resignation returned in March 1940. He had just got over a bout of bronchitis and commented to Campbell:

> "The War is dragging along quite slowly, and I fear it will be a very long business, out of which Europe will come exhausted and impoverished" (80).

Germany invaded Belgium, Holland and Luxembourg on 10th May 1940 and Belgium surrendered on 27th May. De Roy remained in Belgium during "the first days of May", but then took refuge in France. He spent July to September in Toulouse, France, "together with many other Belgians" (81). There he continued his variable star observations with the local society's 4 inch (10 cm) refractor, noting "the sky over there was perfect, with a grand view of the Sagittarius cloud". It is likely that de Roy would have met up with Julien Péridier, who had sponsored de Roy's BAA membership application, as Péridier's La Houga Observatory was not far from Toulouse, but I can find no direct evidence of this. However, France surrendered and was occupied. On returning home to Antwerp on 30$^{th}$ September 1940, he found his house had been commandeered and he had to find temporary accommodation. On a personal note he said:

> "My health is good despite the bitter experiences of the last few months, and rationing is on a sound basis. My eldest son and my son-in-law are now here with us, but my youngest son is still a prisoner of war in Germany" (81).

De Roy refuse to publish his newspaper during the occupation as it would have meant being put under the censorship of the Nazi authorities. With the German occupation in force, mail to foreign destinations was being scrutinised. Part of de Roy's letter to Campbell on the 1$^{st}$ February 1941 has been censored (82):

> "Life is becoming very dreary and difficult over here. [Next one and a half lines of the letter are redacted, with dark ink]. I wonder what is going to happen to us if the War is not finished before next winter. I am still without work, and my youngest son is still a prisoner in Germany, where he is working as a navvy".

He goes on to comment upon the effect that the conditions set out in the Treaty of Versailles, following the First World War, might have had on the current global situation:

> "The Great Powers bare a frightful responsibility in all the present suffering, loss of life and property, and general degeneration of civilised life, for having refused or neglected to see the facts in 1919"

By this time he had moved into a new flat in Antwerp. Because there was no garden, he had to make arrangements to set his telescope up at a friend's house not too far away. The main problem was that there was a curfew between 11 pm and 5 am local time, which curtailed his observations if he didn't wish to stay overnight at his friend's. Although there was a postal service between Belgium and the still neutral USA, he was out of touch with friends in Great Britain and often asked Campbell for news and to convey his best wishes to the BAA-VSS Director and others (82).

The letter to Campbell which accompanied de Roy's AAVSO observations for February 1940 contained some good news (83). His youngest son had at last been freed after 10 months of captivity in Germany and had arrived home in Antwerp. He was severely malnourished, having lost 3 stones and endured a bout of typhoid fever and bronchitis. But in Antwerp "the food situation is still very difficult. We have had no meat now since many weeks".

In May 1941, the "food situation is still very disquieting" and de Roy told Campbell that his April AAVSO submission would be the last for a few months as it was getting darker later, as spring progressed, and the curfew was still in force (84). He was able to resume observing in the second half of September 1941 and he re-silvered his 8 inch (20 cm) mirror in preparation for the new observing season using a small quantity of silver nitrate left over from pre-war days (85). He commented positively that a number of astronomical publications had been received at the Royal Observatory in Brussels, through the good offices of Harvard College Observatory. Bart Bok (1906-1983) had arranged for various astronomical journeys to be sent from the Observatory to various countries in occupied Europe (86) (87).Clearly mail was still passing between Belgium and the USA, although the typical journey time was 6 weeks. Again de Roy referred to the difficult food situation, but about 9 lines of text, presumably about living conditions, had been redacted by the censor. De Roy was able to maintain correspondence with and number of astronomers in occupied Europe. These included Jan Oort in the Netherlands, who had discovered the galactic halo and was later a pioneer radio astronomer, (1900-1992) and Tadeusz Banachiewicz (1882-1954) in Poland, who was Vice President of the IAU during much of the 1930s, and several astronomers at the Meudon Observatory in France. In addition to Campbell, he also corresponded with other US Astronomers including Harlow Shapley at Harvard College Observatory.

De Roy's last letter to Campbell was dated 3$^{rd}$ November 1941 (88) and was again highly redacted. He submitted his October observations (the last being T And on 24$^{th}$ October 1941) and again asked for news of his friends in Great Britain, especially members of the BAA-VSS.

Campbell wrote to de Roy on 3$^{rd}$ December 1941, with some news snippets, for example about a new SS Cyg-type dwarf nova, the recent fall meeting of the AAVSO, and celebrations in honour of Harlow Shapley's 20 years as Director of Harvard College Observatory (89). Clearly Campbell and de Roy had grown close, Campbell sharing news about his grandchildren. No doubt Campbell intended all these comments to be encouraging and to boost de Roy's spirits. Campbell noted :

> "There is not much which I can say about conditions over here. I often try to picture conditions over your way, but not too vividly. We have to listen to reports, weigh the pros and cons, and then draw our own conclusions. I hope things will eventually work out for the best, whatever that may be. We think we know what is best, but we will have to await results".

But de Roy never received the letter since it was returned to AAVSO headquarters marked "This communication returned to sender because it is addressed to an enemy or enemy-occupied country". Pearl Harbour was bombed on 7th December 1941 and the USA had entered the war.

During the winter of 1941/1942, de Roy caught a cold whilst observing variables (2), which eventually turned to bronchitis and pneumonia from which he never recovered. He passed away on 15th May 1942, aged 58. The funeral announcement is shown in Figure 11. One can speculate to what extent poor nutrition may have contributed to his death. Dorrit Hoffleit commented "It is particularly disheartening to realise that starvation in a war-torn country had much to do with the loss of an able man in the prime of life" (90).

A number of speeches were given at the funeral, including one by M.E. Delporte, Director of the Royal Observatory at Brussels. A warm appreciation of his life was written in a special edition of Le Matin on 4/5th November 1944, the first edition after the liberation of Belgium in which tributes were paid to the fallen, which ended:

> "Let us remember the familiar silhouette, the back slightly curved, the lip holding a stump of burned out cigarette, the ironic glare, the aristocratic figure, the nervous hand, the rebellious head of hair" (91)

**De Roy's own variable star observations**

De Roy was a prolific and indefatigable variable star observer. As a newspaper editor he often did not finish work until late at night and then started observing on his return home. It was said that typically a 5 hour rest from 5 am to 10 am would be sufficient (92). In an analysis by Cox and Moreau (93) published shortly after his death, they estimate that he made some 91,000 estimates. The majority of targets were LPVs, reflecting the interests of the main variable star organisations at the time. In addition there were 10 novae, including V603 Aql (443 observations from 1918-1940), DQ Her (520, 1934-1940) and CP Lac (207, 1936-1938). There were also several dwarf novae including SS Aur (363 observations), SS Cyg (2922), and U Gem (1029).

He contributed observations to the AAVSO, the Association Française des Observateurs d'Étoiles Variables (AFOEV) and of course the BAA-VSS. The BAA-VSS computer database contains 6565 observations made between 1906-1908 and 1931-1941, but many observations have yet to be entered. Many observations were published in a large number of papers some of which have already been referred to in this paper, including the BAA-VSS reports and the BAA Memoir. He also wrote review papers on specific stars, including RT Lac (Figure 12) (94), the brightening of $\gamma$ Cas in 1936 (95), and Z And (96).

**Other astronomical observations**

Although de Roy is best known as a variable star observer, he actually started in astronomy by observing meteors (2). He published numerous accounts of his meteor work in the English Mechanic, including a thorough analysis of his observations of the 1907 Aquarids (97). He reported on 3 fireballs "brighter than Venus" that had been seen in the skies over Belgium in May 1908 (98). On 17th October 1917 he observed a fireball with brightness "at least 5 times Venus at her maximum" in Draco, which left a trail that persisted for at least 13 minutes (shown in Figure 13) and which showed structure when observed with his telescope (99). A further fireball was observed on 21st May 1922 which "moved slowly among the stars of Leo and left a tail of sparks like a rocket" (100). On several occasions he attempted accurately to record the paths of meteors across the sky in an attempt to triangulate with other observers (101). At the 1938 Stockholm meeting of the IAU he became President of the IAUs Meteor Commission (as well as becoming Secretary of Commission 27 on Variable Stars).

De Roy also followed comets as they moved across the night sky, including two relatively bright comets in close succession: Comet Cunningham 1940c in December 1940/January 1941 (102) and Comet Paraskevopoulos 1941c (103). He observed the 1910 apparition of Halley's comet over many nights and attempted to point out the comet to people in Antwerp. Whilst he could easily see the tail, his audience were often far less impressed:

> "I, however, had an opportunity to show the comet to other people, who, though being, in the full significance of this word, "profanes," saw it at first sight, and suspected the short tail. Of course, they gave full expression to their disappointment as to the modest character of this "spectacle." To English readers not acquainted with Continental events this may seem rather strange; but it is a matter of fact. The real signification of the astronomical event we now have the pleasure to witness has been completely and very badly denaturated [*sic*] in the public's spirit here, under the influence of a small band of pseudo-scientists"  (104)

These comments were in connexion with the passage of the Earth through the comet's tail, which had received considerable press attention world-wide. He went on to say:

> "As soon as the fact of the probable transit of the Earth through the tail had been announced, a dreadful story of "poisoned cyanide gases" went on, and expanded with a wonderful ease. This "fin du monde" story had found very firm believers, and—though no suicides were coupled with it in this country, as elsewhere—a distinct uneasiness was noticeable on the 18th [May 1910]. But what must one think of the people who on that date spent the night outdoors in order to "see the comet," and afterwards charged "the astronomers" with falsehood?" (104)

In 1932 de Roy travelled to the USA where he was hosted at Harvard College Observatory which was also the headquarters of the AAVSO at the time. There he was able to meet many of his long term correspondents including Leon Campbell and Harlow Shapley. During the visit de Roy was able to achieve a "world first": observing a total solar eclipse from a helicopter (105). The flight took place on 31st August 1932 over Old Orchard, Maine. The helicopter flew to an altitude of 4000 ft and slowly descended to 2,200 ft during totality, thereby taking him above a layer of haze. To increase the sensitivity of his eyes, he kept them closed for 10 minutes before totality. He used a calibrated cross staff to measure as accurately as he could the extent of the corona at totality, concluding that the longest coronal streamer extended to more than four solar diameters. He stayed on in the USA to attend the 4$^{th}$ General Assembly of the IAU in Cambridge Massachusetts in early September. There were two representatives from Belgium: de Roy and the well –known astrophysicist and priest Monsignor George Lemaître (1894-1966; Figure 14), who had proposed what became known as the "Bing Bang" theory of the origin of the universe, which he called his "hypothesis of the primeval atom" (106). One presumes that de Roy met Lemaître in Belgium on other occasions since they were both prominent in Belgian scientific circles. The meeting was also attended by de Roy's successor as VSS Director, W.M. Lindley and his wife. A photograph of de Roy with some of the IAU delegates in shown in Figure 15.

De Roy didn't neglect other astronomical subjects entirely. His publications showed that he observed an eclipse of the moon (107), Jupiter (108), sunspots (108) and the asteroid Eunomia (although he apparently only observed this because it was moving through the field of R Peg in August 1916) (109). He observed the occultation of Aldebaran on 21$^{st}$ Dec 1904 with a 2 inch (5 cm) Vinot refractor, noting "and the star disappeared very suddenly on the dark and invisible level of the satellite, without any projection on the limb, as sometimes observed" and noted "reapparition [*sic*]........without any prominent particularity" (108). He also wrote a paper on the long-disputed secular variation in the colour of Sirius, which reviewed the historical evidence (110). He carried out his own estimates of the brightness of Sirius (relative to Rigel, Procyon and Aldebaran). Thirty-two separate estimates over 6 nights gave a mean magnitude of -1.4, with a standard deviation of 0.3 magnitudes.

As an example of de Roy's broader interests, he also remarked on the presence of *floaters* in his eyes (111): dark objects moving across the field of view, sometimes caused by detached cells from the retina moving within the eye's *vitreous humour*. He "noticed in 1903 and 1904 a dark dot which was seen best at morning at the first admission of light in the right eye. At the end of 1904 I ceased to see it. About the middle of 1905, however, it was substituted by a strange object that I could not identify at the beginning. I recognised later on that it was a sort of crescent with rounded horns divided in the side of the breadth in three parts by two partitions. I opine this object is an aggregate of three cells detached from an inner vessel of the eye; but as I am not acquainted with animal microscopy, I cannot say more about it".

In addition to astronomy, de Roy was also interested in meteorology, drawing attention to "monster hailstorm" in Spa, southern Belgium on 17$^{th}$ June 1904 (112).

Many hailstones were "as great as hen's eggs", 10 cm (4 inch) diameter and 100 grammes (4 oz) in weight fell, with one being 320 grammes (11.5 oz). Windows and even tiles were broken over a wide region. He reported on a further hailstorm in the Verviers region of southern Belgium on 14th August 1914, which, in addition to causing damage to property, killed many birds.

**Honours and recognition**

De Roy's contribution to astronomy was widely recognised in Belgium and internationally. He served on many important bodies such as IAU commissions on Variable Stars and on Meteors. He was selected as a member of Belgium's Comité National D'Astronomie when it was founded in 1919 (he was especially active in working through this body to encourage international collaboration with other in variable star organisations (113)). We have noted that he was an honorary member of the AAVSO and he received many other honours. In 1936 he received the honorary degree of Doctor of Mathematics and Physics by the University of Utrecht, the Netherlands, on the occasion of the tercentenary of that university (114). De Roy had had a long association with Prof. A. A. Nijland (1868 – 1936; Figure 16) of Utrecht University, a well-known variable star researcher, for many years (115). A 43 km lunar impact crater was named for him that is located on the far side of the Moon, just behind the south-western limb.

**Conclusion**

De Roy will be best remembered for his contribution to variable star astronomy over many years, which brought international recognition. Not only did he contribute through making his own observations, but also by encouraging others, be they fellow Belgians, members of the BAA, or other organisations, to do likewise. He did this by giving presentations at meetings and by publishing many articles which explained how to go about making observations and how to improve one's technique. He wasn't merely content with amassing data, but he took a particular pleasure in analysing them and making them available to a wider community by writing numerous papers. In this way, hard-won observations contributed by many observers didn't merely remain hidden in some archive or database, but were used to advance our knowledge of variable stars. The resulting publications are of lasting value and a tribute to de Roy's professionalism.

**Acknowledgements**

A great many people have assisted me in preparing this paper. I am especially indebted to Tom Williams of Houston (AAVSO) who freely shared with me a large number of papers he had obtained during his research into de Roy which led to the paper he presented at the Brussels AAVSO meeting in 1990 (4). Tom also provided helpful comments on a draft of this paper. In turn, Tom Williams had received much help, especially in regard to locating some more obscure references and in translating some key texts, all of which were made available to me, by Willy de Kort,

## Address

"Pemberton", School Lane, Bunbury, Tarporley, Cheshire, CW6 9NR, UK [bunburyobservatory@hotmail.com]

| Title of VSS report | Reference |
|---|---|
| LPVs in 1921 (1) | (116) |
| LPVs in 1921 (2) | (117) |
| "Irregular" Variables in 1921 and 1922 [a] | (118) |
| LPVs in 1922 (1) | (119) |
| LPVs in 1922 (2) | (120) |
| LPVs in 1923 (1) | (121) |
| LPVs in 1923 (2) | (122) |
| "Irregular" Variables in 1923 and 1924 [a] | (123) |
| LPVs in 1924 (1) | (124) |
| LPVs in 1924 (2) | (125) |
| LPVs in 1925 (1) | (126) |
| LPVs in 1925 (2) | (127) |
| LPVs in 1926 (1) | (128) |
| LPVs in 1926 (2) | (129) |
| LPVs in 1927 (1) | (130) |
| LPVs in 1927 (2) | (131) |
| LPVs in 1928 (1) | (132) |
| LPVs in 1928 (2) | (133) |

**Table 1: VSS reports published by de Roy in the 1920s**

[a] Includes SS Aur, R CrB, U Gem, R Sct,

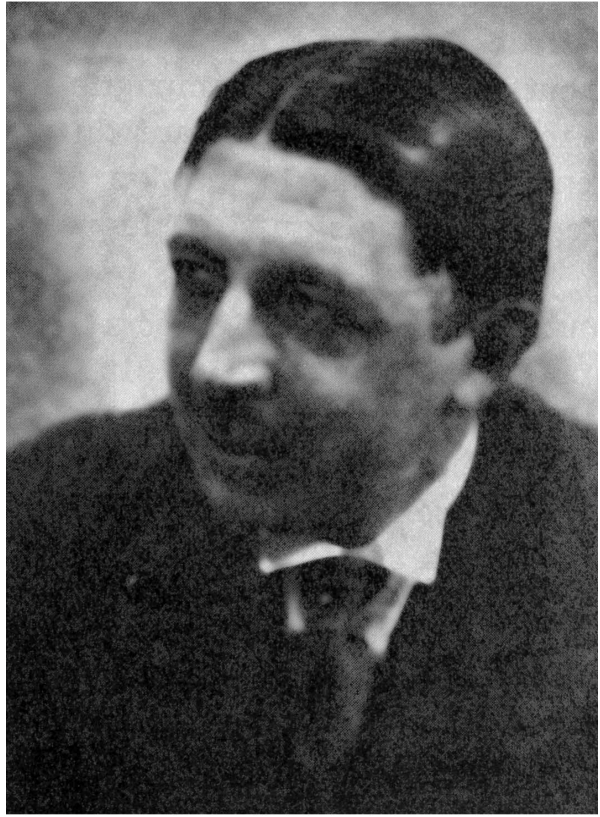

**Figure 1: Portrait of Félix Eugène Marie de Roy**
Date unknown

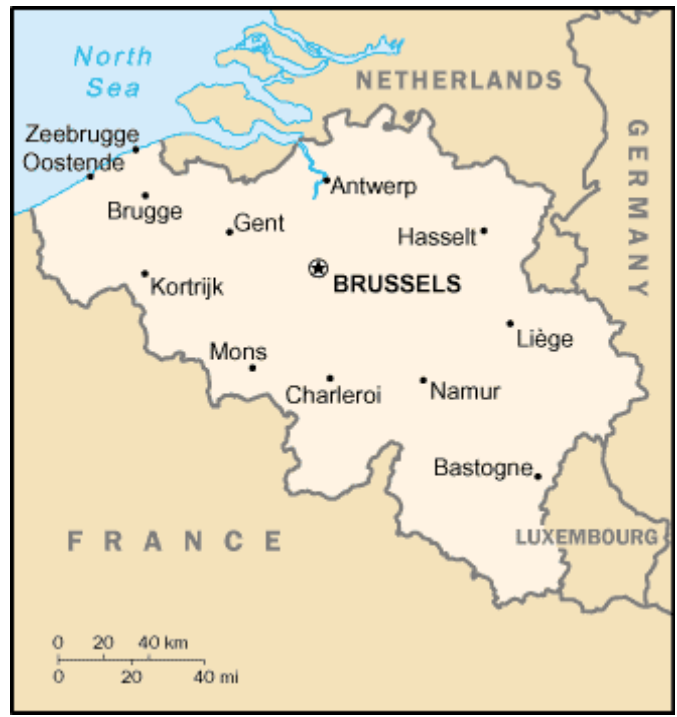

**Figure 2: Map of Belgium**

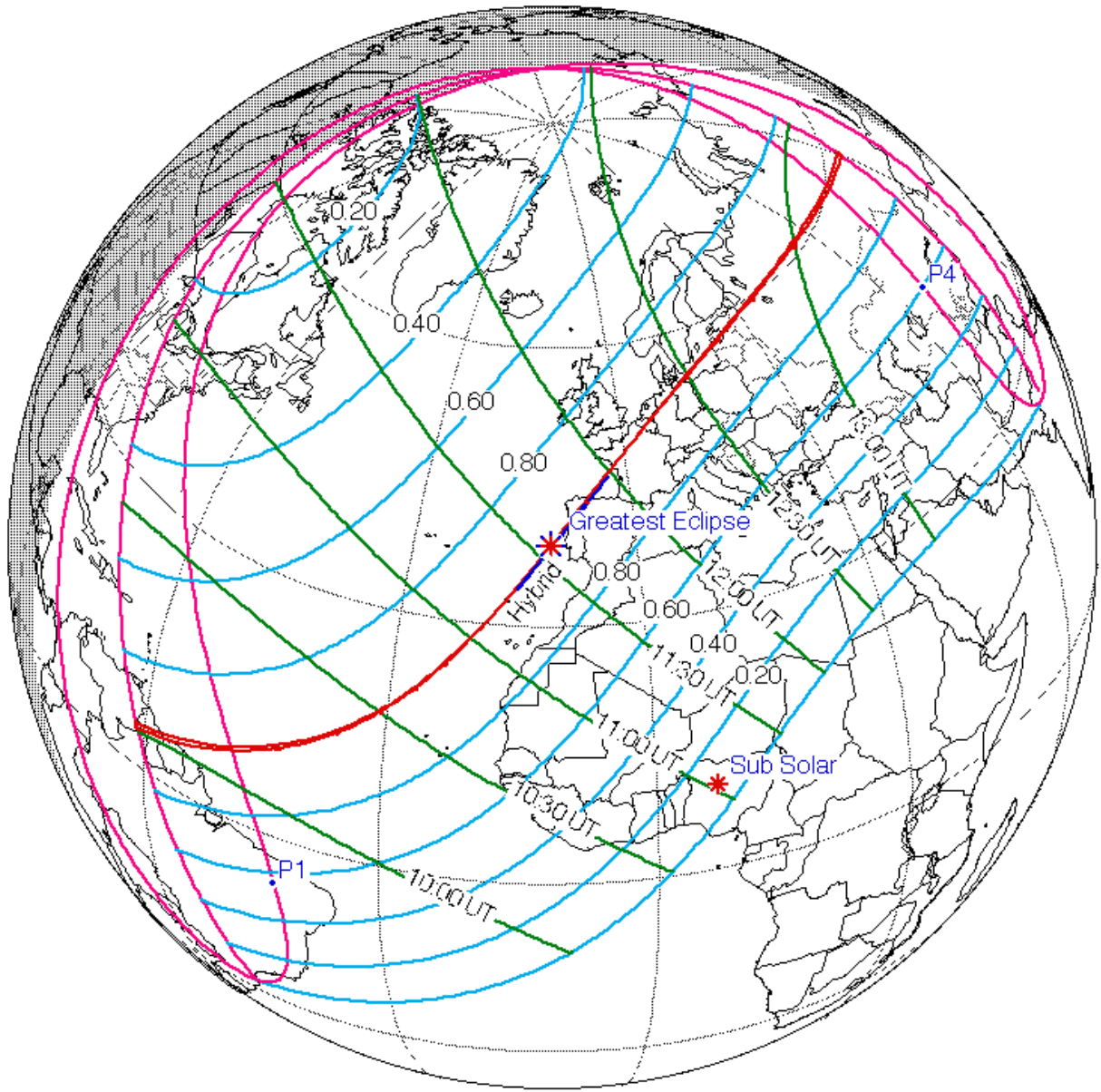

**Figure 3: Path of the solar eclipse of 17th April 1912**

The red line shows the path of totality/annularity, passing from the Atlantic Ocean, through the Iberian Peninsula, France, Belgium, the Netherlands, Germany and on into eastern Europe. Source: NASA

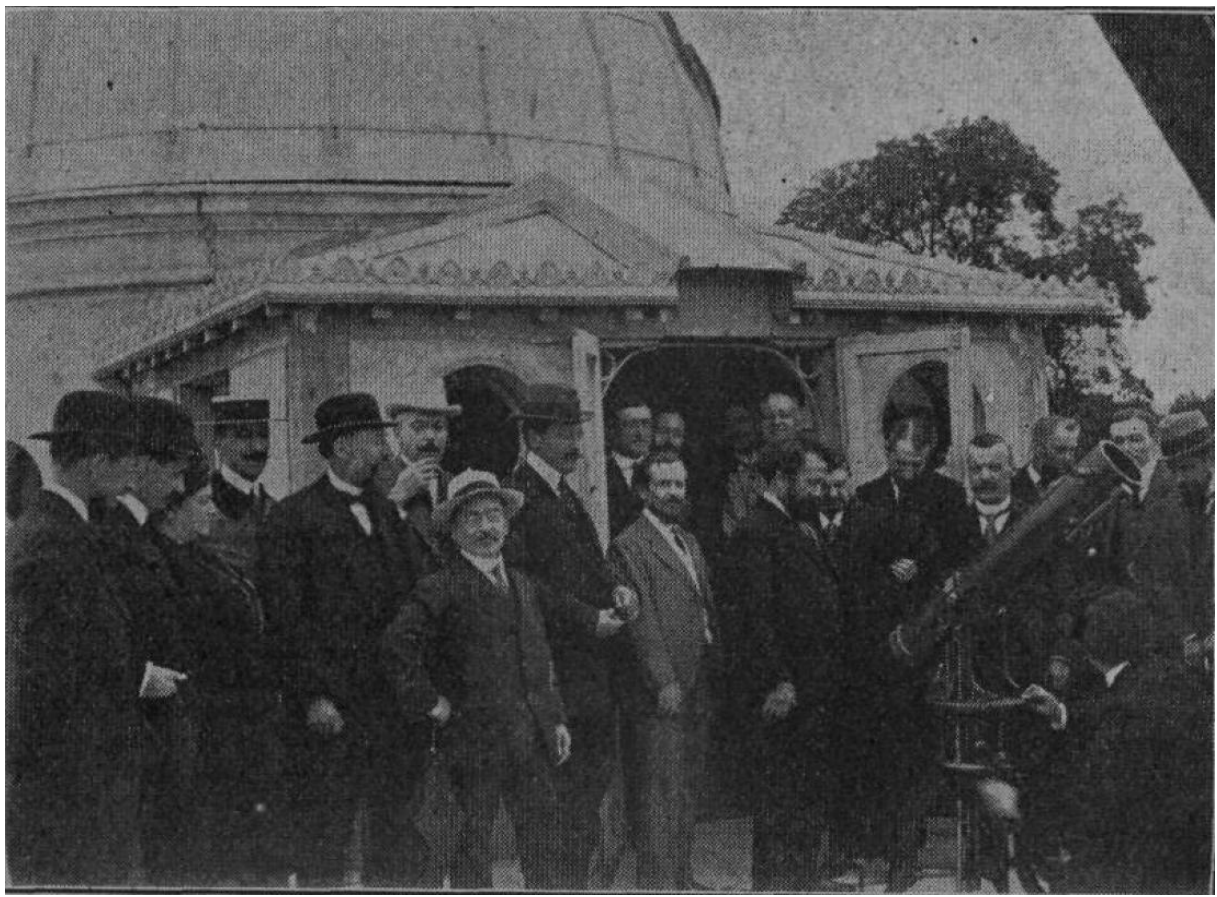

**Figure 4: Some participants of the 1914 astronomical congress visiting Camille Flammarion's observatory at Juvisy**

Flammarion (bearded) is seen in profile standing behind the small telescope. De Roy is the standing in the doorway, third from left, but dimly seen. The gentleman wearing the white felt hat at the front, just left of centre is Don Salvador Eaurich, founder and secretary of the Sociedad Astronomica de Barcelona (14).

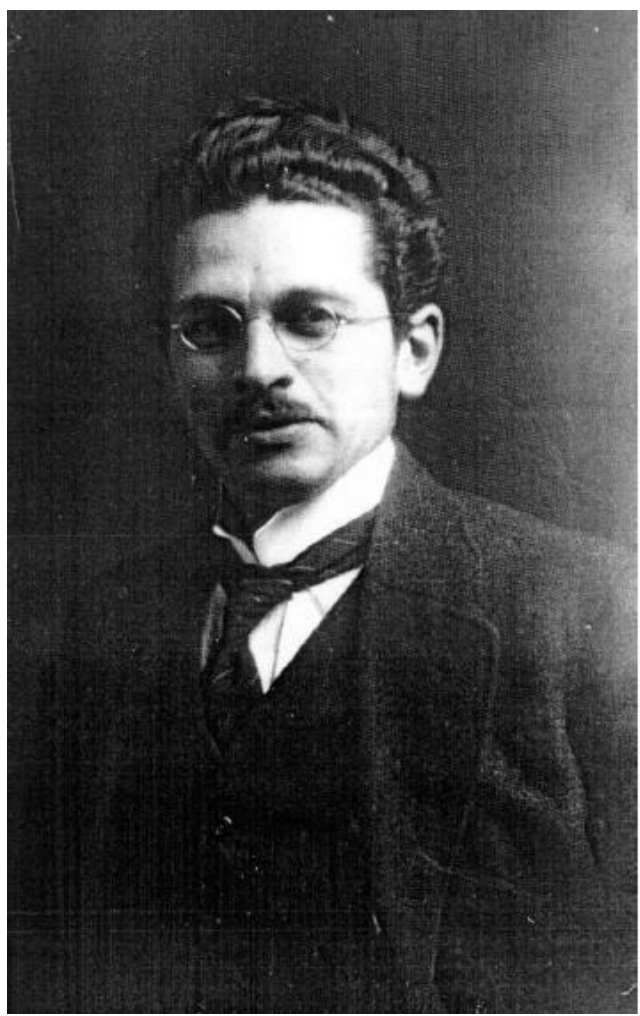

**Figure 5: Robert Jonckheere**

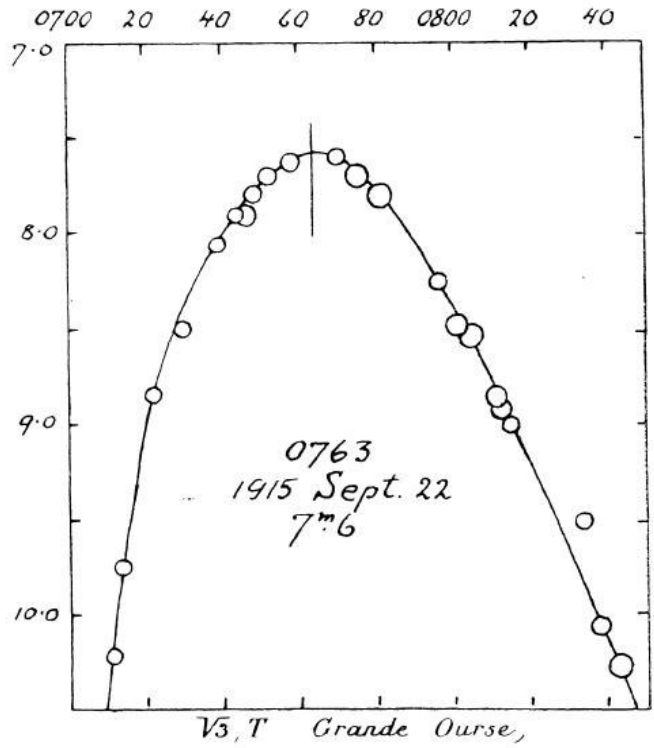

**Figure 6: Light curve of T UMa in 1915** (32)
The x-axis is JD – 2420000, y-axis is visual magnitude

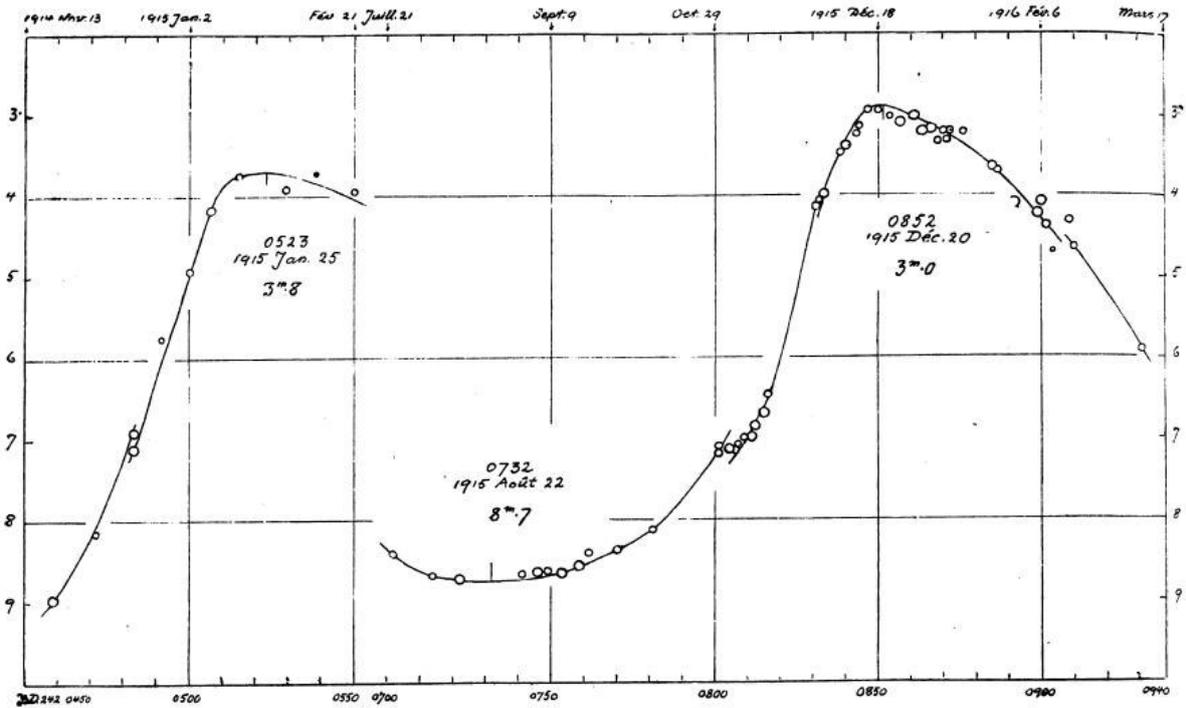

**Figure 7: Light curve of Mira Ceti in 1915** (31)
Times of maxima and minima are marked

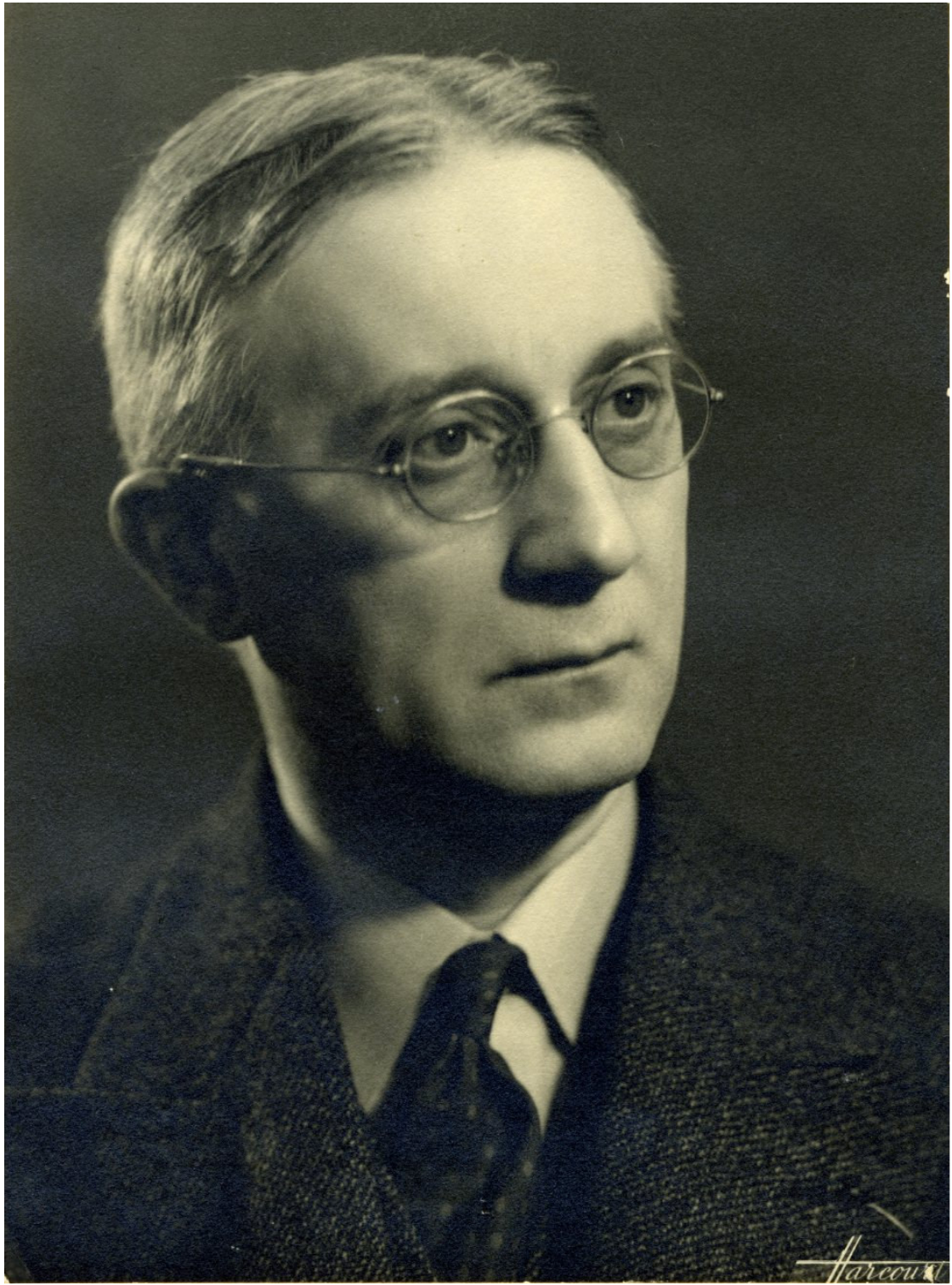

**Figure 8: Julien Péridier**

## British Astronomical Association
### VARIABLE STAR SECTION

---

### CIRCULAR 1.

Deurne, Antwerp, Belgium, 1922 March 31.
23, rue Thisius.

(1) The new Director begs to present his compliments to all the members of the Section. He intends to publish from time to time a circular dealing with various aspects of our work which are of a technical nature, and therefore unsuitable for reproduction in the limited space available in the *Journal*.

He begs to refer to the April number of the *Journal*, which will contain a general statement dealing with his nomination, with the able work of his two distinguished predecessors, with the Sectional programme, which he does not intend to modify for the present, and giving some hints as to the future.

He hopes that all will take the greatest possible interest in our common work, which can only be satisfactorily performed with the active assistance and the goodwill of each individual member.

(2) WORKING LIST OF VARIABLE STARS. — With the addition, at Mr. C. L. BROOK's request, of R Persei, the Working List of the Section now contains the following 54 stars :

A. — SS Cygni.

B. — 21 Long Periods. P < 300 days :

| | | |
|---|---|---|
| * S Aquilæ | R Camelo. | * RY Ophiuchi |
| * R Arietis | * X　　» | * X Pegasi |
| * X Aurigæ | W Coronæ | R Persei |
| R Boötis | * W Cygni | * V Tauri |
| S　　» | R Draconis | S Ursæ Majoris |
| * U　　» | * T Herculis | T　　» |
| V　　» | * W Lyræ | * R Vulpeculæ |

**Figure 9: Front page of BAA VSS Circular number 1 (31$^{st}$ March 1922)**

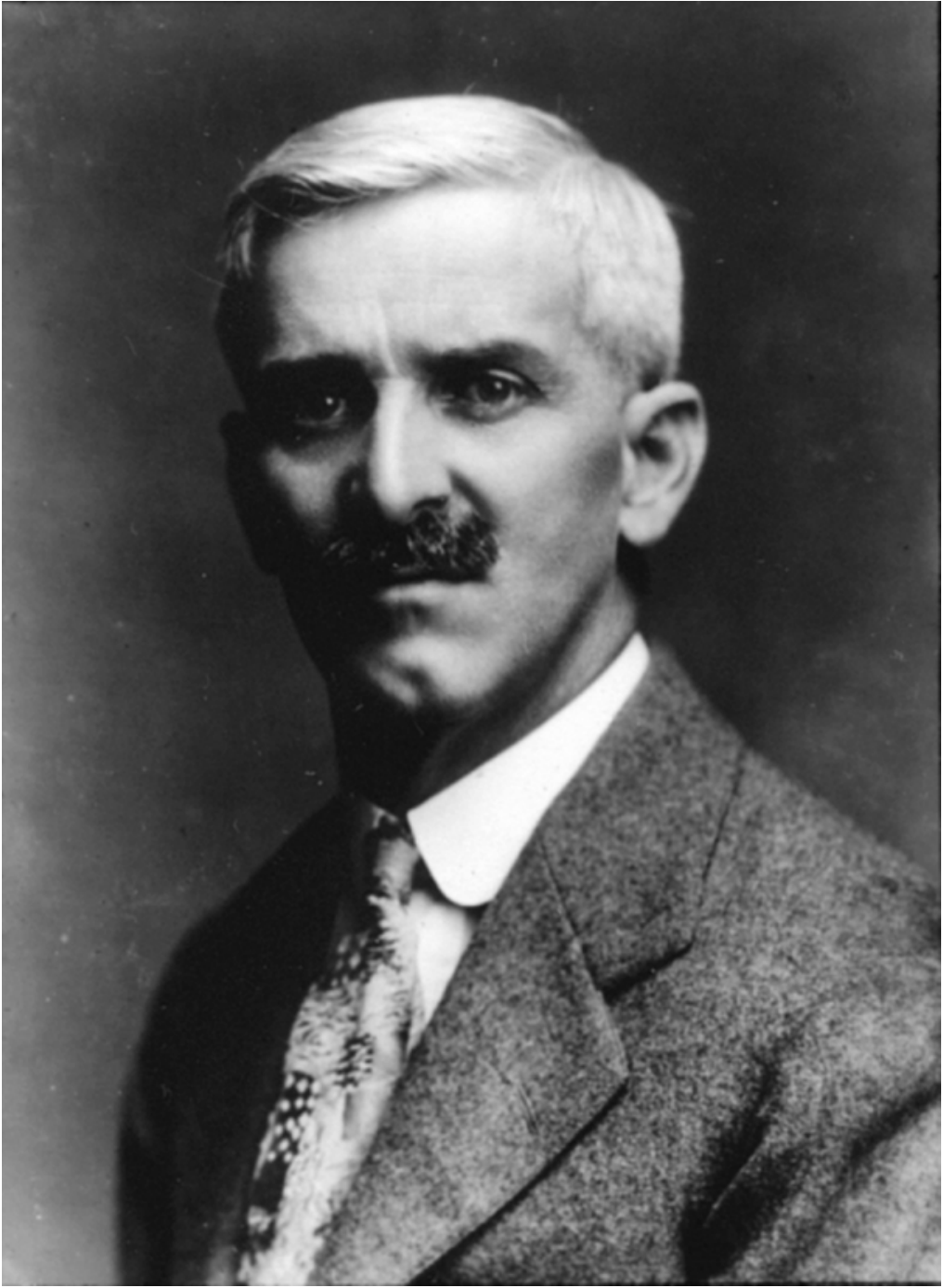

**Figure 10: Leon Campbell**

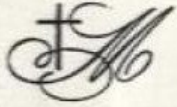

Madame Félix DE ROY, née Louise MEYER ;

Monsieur André SALTI, Madame André SALTI, née Ivy DE ROY ;
Monsieur Georges DE ROY, Madame Georges DE ROY, née Jeanne FANIS ;
Monsieur Jean DE ROY ;
Mademoiselle Estelle DE ROY ;

Monsieur et Madame Roger DE ROY, leurs enfants, beau-fils et petit-fils ; Monsieur et Madame Willy DE ROY ; Monsieur Herman SCHOONHOVEN et Madame, née Germaine DE ROY et leur fille ;

Madame David MEYER, son fils, ses beaux-enfants et petits-enfants ; Monsieur et Madame John DUFOUR ; Madame Charles VERBERT, ses enfants, beau-fils et petite-fille ;

Les Neveux, Nièces, Cousins et Cousines

ont la profonde douleur de vous faire part de la perte cruelle et irréparable qu'ils viennent d'éprouver en la personne de

## Monsieur Félix Eugène Marie DE ROY

Docteur " Honoris Causa " ès Sciences Physiques et Mathématiques de l'Université d'Utrecht,
Rédacteur en Chef du journal " Le Matin,"
Président de la Société d'Astronomie d'Anvers,
Ancien Président de la Fédération des Cercles Scientifiques de Belgique,
Ancien Président de la Section Anvers-Limbourg de l'Association Générale de la Presse Belge,
Fellow of the Royal Astronomical Association,
Chevalier de l'Ordre de Léopold,
Officier de l'Ordre de la Couronne,
Officier de l'Ordre de Wasa,
Chevalier de l'Ordre de St Olaf,

leur époux, père, beau-père, frère, beau-frère, oncle, grand-oncle et cousin bien-aimé, né à Anvers, le 25 juillet 1883, pieusement décédé à Boitsfort le 15 mai 1942, muni des Sacrements de Notre Mère la Sainte Eglise.

Vous êtes prié d'assister au service funèbre, qui sera célébré en l'église paroissiale de St Joseph à Anvers, le mercredi 20 mai à 10 heures.

Réunion à l'église.

L'inhumation aura lieu dans la concession de la famille au cimetière de Vieux-Dieu.

### PRIEZ POUR LUI.

Anvers, le 15 mai 1942.
Avenue de Belgique, 48.

1704

**Figure 11: Announcement of de Roy's death and funeral arrangements** (134)

As is typical of funeral announcements of this era, it ends with "Priez pour lui", "Pray for him"

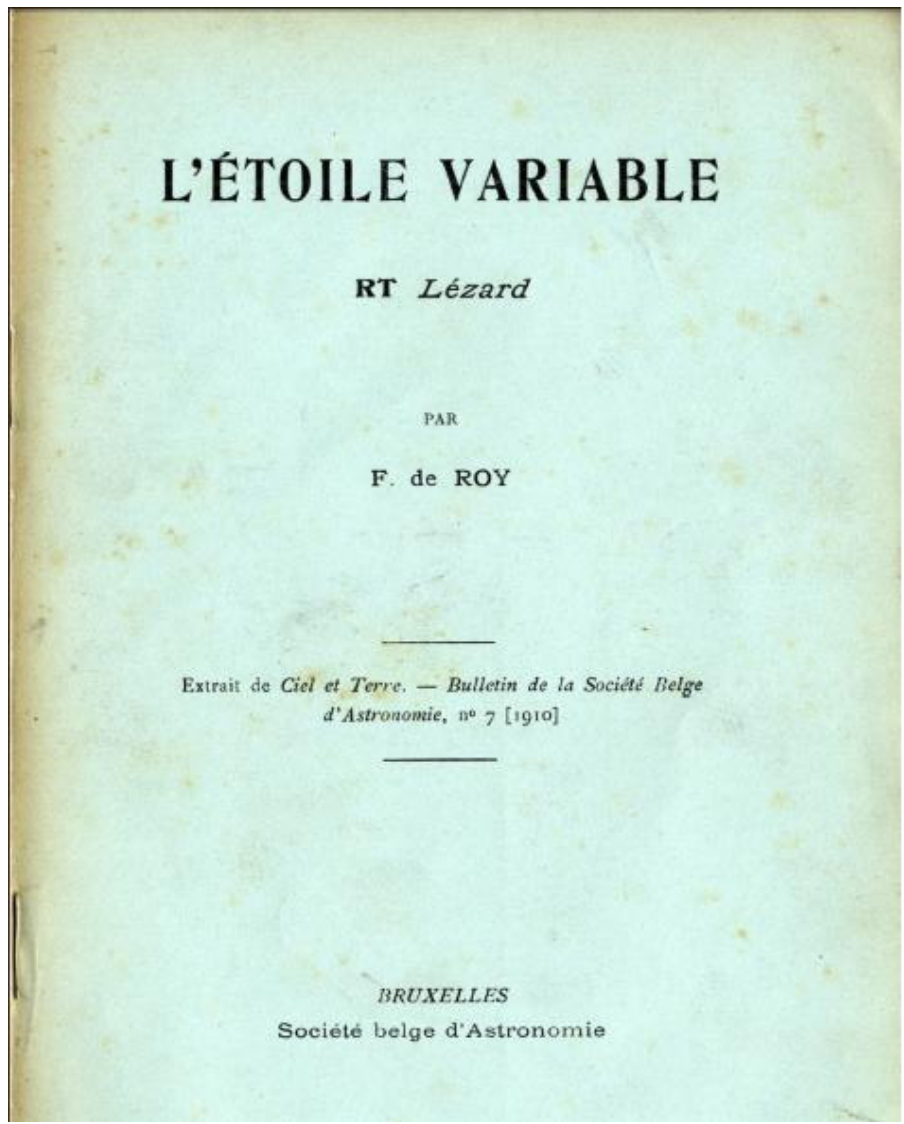

**Figure 12: de Roy's 1910 paper on RT Lacertae** (94)

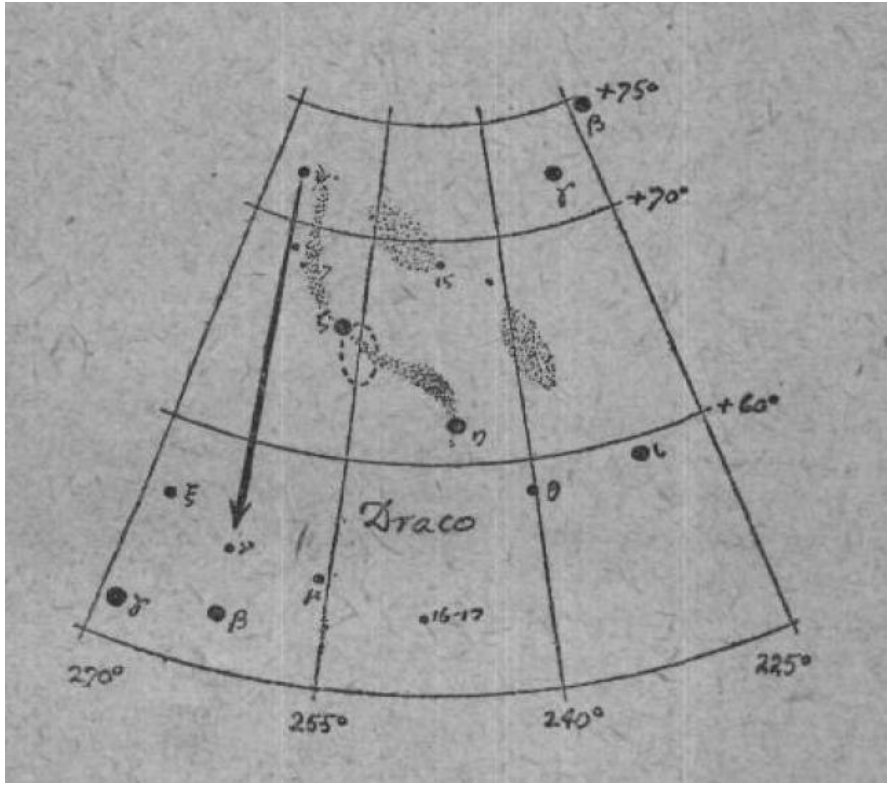

**Figure 13: Path of a fireball seen in Draco on 17th October 1917**
The track of the fireball itself is shown by the arrow. The curved shaded area to the right of the arrow from about +72 to +61 deg of Dec. shows the extent of the trail 5 minutes after the appearance of the fireball; the dotted oval near ζ Dra indicates the brightest region. The two shaded regions near 15 Dra represent the trail 13 minutes after the fireball appeared, at which point the trail was hardly perceptible
(99)

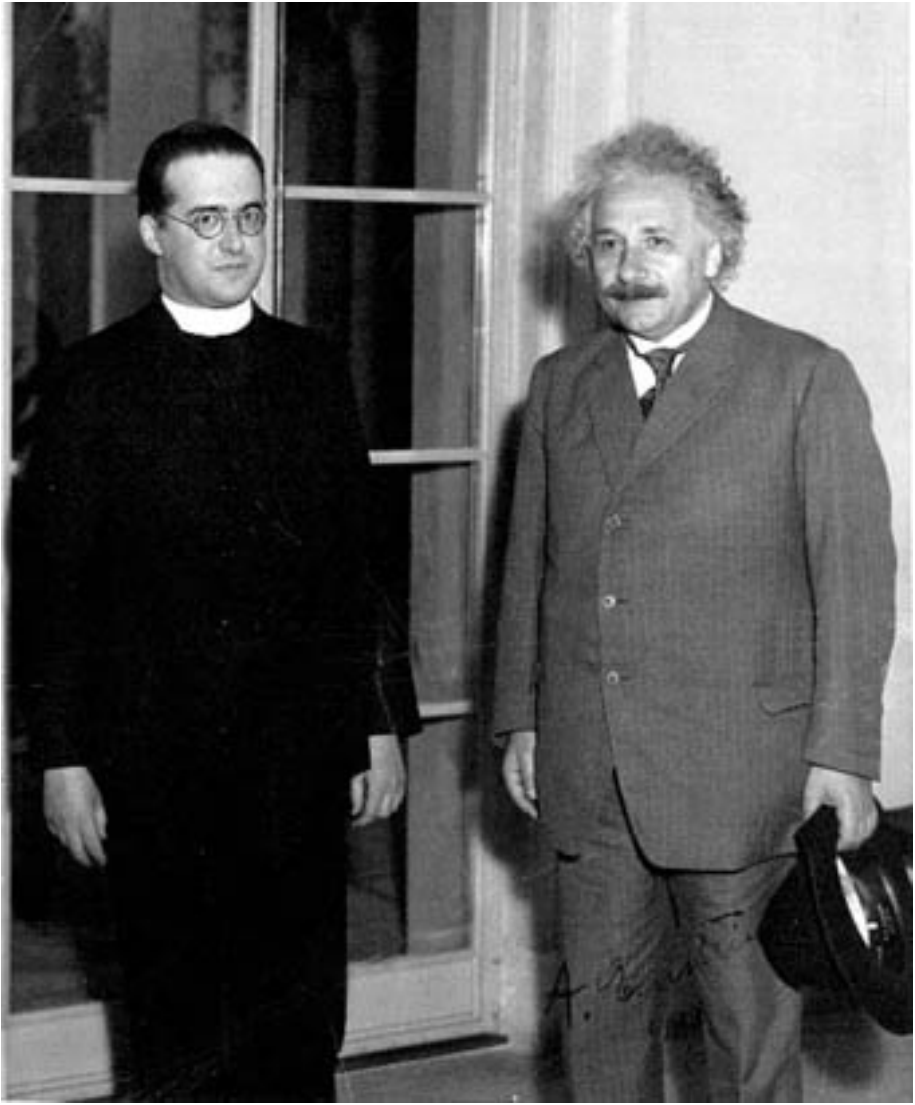

**Figure 14: Monsignor George Lemaître (left) and Albert Einstein**

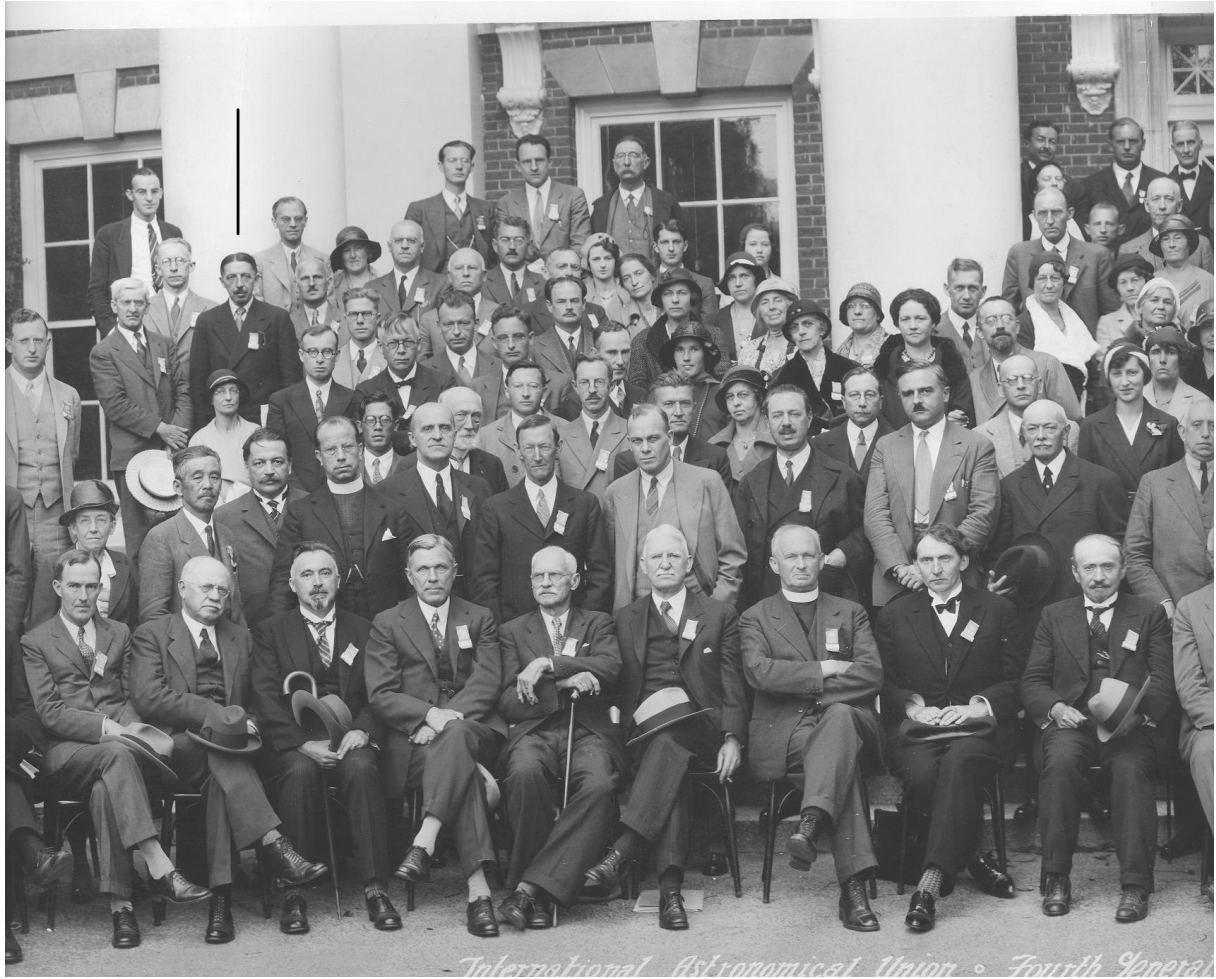

**Figure 15: Delegates at the 1932 IAU meeting in Cambridge, Massachusetts**

De Roy is indicated by the black line. Two to the left of him (as viewed by the reader) is Leon Campbell. Between them is H.L. Alden (1890-1960) of Yale University. The astrophysicist Donald Menzel (1901-1976) is to the left of Campbell. In the back row at the centre of the photograph, the left most of the three men, is Fred Whipple (1906-2004), proposer of the "dirty snowball" hypothesis of comets (135).

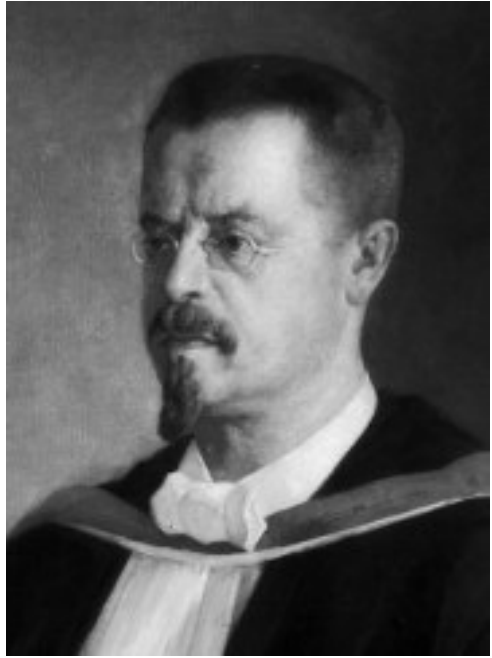

**Figure 16: Prof. A.A. Nijland**